\documentclass[5p]{elsarticle}
\usepackage{lineno,hyperref}
\usepackage{color}
\usepackage[utf8]{inputenc}
\usepackage[T1]{fontenc}
\usepackage{textcomp}
\usepackage{siunitx}
\modulolinenumbers[5]

\journal{Scripta Materialia}

\biboptions{sort&compress}

\begin{document}

\begin{frontmatter}
\title{On the correlation of shear band formation and texture evolution in $\alpha$-brass during accumulative roll bonding}

\author[lww]{Marcus Böhme\corref{cor}}
\ead{marcus.boehme@mb.tu-chemnitz.de}
\author[lww]{Martin F.-X. Wagner}
\ead{martin.wagner@mb.tu-chemnitz.de}
\address[lww]{Chemnitz University of Technology, Institute of Materials Science
	and Engineering, Erfenschlager Str. 73, 09125 Chemnitz, Germany}


\cortext[cor]{Corresponding author}

\begin{abstract}
We  studied the microstructural evolution of the low stacking fault energy $\alpha$-brass alloy CuZn15 during
accumulative roll bonding (ARB). Most notably, the typical brass-type texture was
clearly observed after four ARB passes (approx. 93.8~\% total thickness reduction), before significant shear localization set in. 
This observation contradicts the widely accepted idea that shear band formation is a necessary prerequisite for the development of the brass type texture, indicating that the two phenomena, shear banding and development of the brass texture, are only correlated in ARB, and that their order of appearance can be switched depending on experimental parameters.
\end{abstract}


\end{frontmatter}

It is well-documented that two distinct rolling textures can occur in face-centered cubic (fcc) metals \cite{Hu52,Wassermann63,Duggan78,Hirsch88,Leffers90,Leffers09}. The type of texture developed during cold rolling  strongly correlates with the stacking fault energy (SFE) of the metal or alloy. Materials with a high SFE, like aluminum, develop a so-called copper-type texture with prevalent \{112\}<111> (copper) and \{123\}<634> (S) texture components
\cite{Wassermann63,Duggan78,Hirsch88,Leffers09}. In contrast, low SFE alloys like $\alpha$-brass with medium to high zinc contents predominantly show  \{011\}<211> (brass) and \{011\}<100> (Goss) orientations -- appropriately also called brass-type texture\cite{Leffers90,Leffers09}. Many low SFE materials like Cu-Zn alloys \cite{Hirsch88,Duggan78,Leffers90}, Cu-Al alloys \cite{Paul02,Paul03} or pure silver\cite{Paul03,Paul07} form fine deformation twins during the initial stages of plane strain deformation (i.e., during rolling). The tightly packed twin boundaries inhibit conventional dislocation glide and it is commonly assumed that shear bands are created at large deformations as the only mechanism to accommodate the imposed strains \cite{Hirsch88,Duggan78,Paul02,Paul03,Paul07}. In fact, nucleation and growth of these shear bands are typically considered to be the main mechanism that facilitates the subsequent development of the brass-type texture components \cite{Paul07,ElDanaf00,Jia12,Jia12a}. 

While the correlation (and apparent causality) of shear band formation and brass-type texture evolution has been well documented during conventional rolling, much less is known about their interaction in other processes with a near plane-strain deformation. The accumulative roll bonding (ARB) process, originally proposed by Saito et al. \cite{Saito99}, for instance, is an interesting tool to investigate severe plastic deformation in sheet materials. By repeatedly cutting and stacking the material between individual rolling passes, large amounts of plastic strain can be accumulated and ultrafine-grained (UFG) microstructures can be produced while keeping a constant sheet thickness. Nominally, five ARB passes correspond to a von Mises equivalent plastic strain (determined as defined in \cite{Saito99}) of $\varepsilon_{\mathrm{eq}} = 4.0 $, i.e., to a total thickness reduction of approx. 97~\%. 
One advantage of ARB is that the significant amounts of shear strain (i.e., redundant shear strains that primarily occur near the surface) can lead to faster grain size reduction \cite{Huang03}. 
Compared to conventional rolling, much less has been published on shear banding and texture evolution during ARB, and the literature appears to be partially incomplete. 
For instance, Pasebani et al. observed brass and Goss as the dominant texture components after three ARB cycles in a 70/30 brass; they discussed shear banding as main cause for the texture transition similar to conventional rolling \cite{Pasebani10,Pasebani10a}. Chen et al. also reported behavior similar to conventional rolling in pure copper with a \{211\}<111> copper texture after six ARB cycles \cite{Chen09}. Shaarbaf et al., in contrast, reported the development of a shear type texture in copper after four ARB passes\cite{Shaarbaf09}. 
Finally, to the best of our knowledge only little research has been published on ARB of alloys that can develop either a copper-type or a brass-type texture depending on the experimental conditions, despite the fact that such investigations may shed more light on the interesting topic of the interaction of shear band formation and texture development. In the present work we therefore study microstructural (via scanning electron microscopy, SEM) and texture evolution (via X-ray diffraction, XRD) of an $\alpha$-brass with 15~wt-\% zinc during ARB. 
This allows for a comparison to well-established results obtained by Leffers et al. for conventional rolling with intermediate thickness reductions \cite{Leffers88, Leffers90, Leffers93}. 
CuZn15 is of special interest as it shows a dominant brass-type texture after cold rolling although the overall volume of twins formed is only a few percent \cite{Leffers90, Christoffersen97}.

A commercially available CuZn15 sheet material (obtained from Mecu GmbH, Velbert, Germany) 
with an initial thickness of 2~mm was cut to strips (100~mm x 30~mm), rolled down to a thickness of 1~mm and recrystallized at 450~\textdegree C for 30~min. Average grain size after recrystallization was $ 12 \pm 4 \,\mu m$. 
ARB was performed on a laboratory rolling
mill with 190~mm roll diameter at a  rolling speed of {0.06~m~s\textsuperscript{\={1}}} without lubrication. The stacked sheet samples were pre-heated to 473~K before rolling to improve the bonding quality. After each ARB pass, a 1~cm long strip was cut from the center of the samples for SEM analysis and XRD texture measurements. The SEM samples were cut in the RD-ND plane, ground and polished with final polishing on a vibratory polisher (Buehler VibroMet 2) using 50~nm alumina suspension. SEM investigations were performed
 with a Zeiss NEON 40 EsB field emitting SEM equipped with a retractable four
quadrant backscattered electron detector (QBSD) mounted under the pole piece. 

\begin{figure*}
\centering
	\includegraphics[scale=0.75]{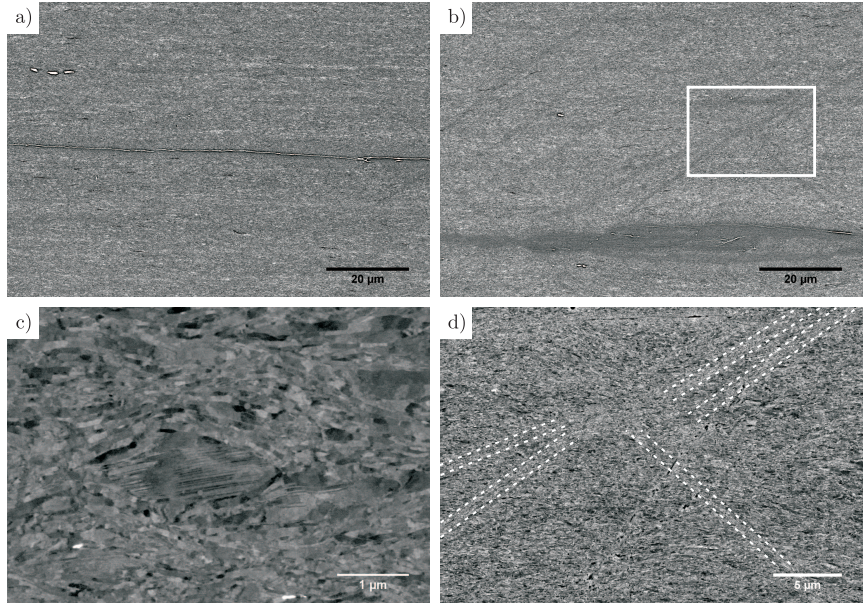}
\caption{QBSD electron micrographs taken from the RD-ND plane between ARB passes. The top row shows overview micrographs after (a) 4 and (b) 5 passes. The contrast in these micrographs was inverted to increase the visibility of bonding zones and shear bands. Higher magnification micrographs after (c) 3 and (d) 5 ARB passes are shown in the bottom row without contrast inversion. The white frame in (b) marks the position of the detailed view in (d). Dashed white lines in (d) indicate the orientations of several distinct shear bands. }
\label{fig:SEM}
\end{figure*}

The SEM investigations (Figure~\ref{fig:SEM}) show that the material generally exhibits the well-known microstructural deformation mechanisms associated with ARB: we observed both a reduction of grains size with increasing number of passes and mechanical twinning. After four ARB passes ($\varepsilon_\mathrm{eq} = 3.2$ or 93.8~\% total thickness reduction), the microstructure is macroscopically homogenous along the rolling direction (figure~\ref{fig:SEM}.a). Heterogeneities/ microstructural gradients in the normal direction are a direct result of ARB processing. The distinctive stripe in the  center of the micrograph marks the position of the bonding zone that was formed during the fourth ARB pass. The two symmetrically arranged stripes above and below originate from previous ARB passes. Additional inhomogeneities are observed at a much smaller length scale (figure~\ref{fig:SEM}.c) up to three ARB passes. Surrounded by grains that have been refined already to an average subgrain diameter well below 500~nm, there are some significantly larger, residual grains, containing densely packed twins with twin boundaries oriented parallel to the rolling direction (clearly visible as striped patterns in the QBSD micrographs; the twins are expected to be of type \{111\}<112>). These microstructural observations are consistent with the low amount of mechanical twins reported by Leffers and Bilde-S{\o}rensen \cite{Leffers90} and Christoffersen and Leffers \cite{Christoffersen97} after conventional rolling. It is likely that twinning contributes to stabilization of the few relatively large grains.

It is important to highlight that no shear bands are observed after 4 ARB passes, whereas after the fifth pass ($\varepsilon_\mathrm{eq} = 4.0$ or 96.9~\% total reduction), many shear bands that run diagonally through the RD-ND plane with two distinct orientations are visible in the QBSD micrographs as dark stripes (figure~\ref{fig:SEM}.b). The micrograph at higher magnification (figure~\ref{fig:SEM}.d) reveals the complex interaction of the intersecting shear bands. Most interestingly, the two shear bands running from the top left to the bottom right are aligned parallel to each other prior to their intersection with the band running from the top left to the bottom right. After the intersection, the lower shear band is offset with respect to its original alignment. The upper shear band is also offset and in addition it also exhibits a changed orientation towards the rolling direction. This demonstrates that, although all shear bands were formed during the fifth ARB pass, shear band formation is spread out over a certain time period. While this is beyond the scope of the present work, careful microstructural analysis of the geometric offsets at intersection points allows in principle to analyze the formation history and sequence of individual shear bands. No information has been published, for either conventional rolling or for ARB, on the formation of shear bands in the CuZn15 alloy investigated here. Pasebani and Toroghinejad reported shear band formation in 70/30 brass already after two ARB passes \cite{Pasebani10a}. The later onset of shear band formation in the present work most likely can be explained by the higher stacking fault energy \cite{Grace70}, and consequently by a lower twin density. 

To further investigate the influence of the relatively late formation of shear bands on texture evolution, we performed complementary XRD texture measurements. Prior to these, samples were ground to a depth of approximately 400~$\mu m$ in the RD-TD plane to reduce the influence of redundant
shear strains near the surface. Textures were investigated by measuring the \{111\}, \{200\} and \{220\} pole figures of the samples with a Siemens D5000 diffractometer in reflection geometry using Cu-$K_\alpha$ radiation. 
Crystallographic orientation distribution functions (ODFs) were calculated from these pole figures using the {MTEX} toolbox for MatLab \cite{bachmann10}. 
\begin{figure}
	\centering
		\includegraphics[scale=0.75]{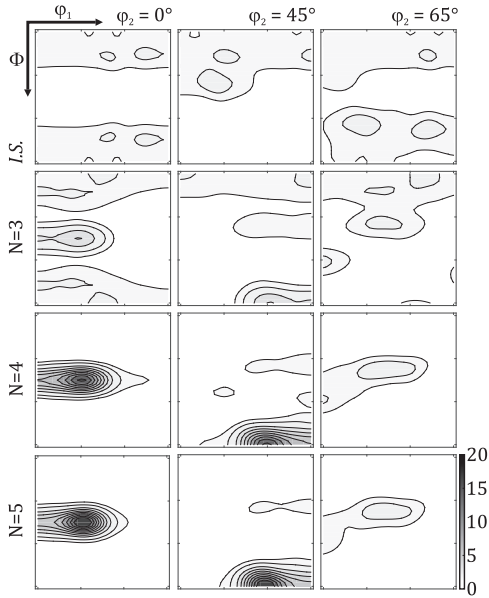}
	\caption{Orientation distribution function (ODF) sections determined by XRD at constant
angles $\varphi_2$ (0\textdegree, 45\textdegree and 65\textdegree) of the brass sheets in the initial state (I.S.) and after 3,4 and 5 ARB cycles, respecitvely. While the recrystallized sheets do not exhibit any significant rolling textures, brass and Goss components are visible after 3 ARB passes together with weak copper and S components. In subsequent ARB passes, the brass and Goss components become dominant with further declining contributions from copper and S components.  
}
	\label{fig:CuZn15_texture}
\end{figure}

Figure~\ref{fig:CuZn15_texture} shows ODF sections at constant angle $\varphi_2$ similar to the data representation in Jia et al.\cite{Jia12} (Euler angles written in Bunge notation \cite{Bunge69}). The sections at $\varphi_2 = \ang{45}$ (central column in fig.~\ref{fig:CuZn15_texture}) exhibit a lack of shear type texture components in all samples, except for a minor peak around \{111\}<211> at N=4. 
Clearly, shear strains from previous ARB passes hardly affect the global texture evolution. 
The ODF of the initial, recrystallized state (I.S.) 
suggests the existence of a weak texture with no significant contribution from either copper or brass type components. Already at N=3 ($\varepsilon_\mathrm{eq} = 2.4$ or 87.5~\% total thickness reduction) the brass and Goss components are clearly visible, together with weak contributions from the copper and S components. After four ARB passes, the brass and Goss components have become the most dominant components while the copper component contribution is strongly reduced. The texture after the fifth ARB pass is almost identical to the previous one, but exhibiting a continued reduction of the copper component.

At first glance, the texture evolution documented in fig.~\ref{fig:CuZn15_texture} is consistent and in line with previously established models (e.g. \cite{Paul03, Jia12}): With increasing degree of deformation the low SFE material develops a dominant brass-type texture; the plastic strains that are necessary to produce a dominant brass type texture are somewhat higher than during conventional rolling \cite{Leffers88,Leffers93}. There is, however, a key difference to previous reports in terms of the sequence of shear banding and texture evolution: A fully developed brass type texture is observed at $N=4$, but shear bands (which are often considered as a necessary prerequisite for brass type texture formation) only occur later (i.e., in the QBSD micrographs in fig. ~\ref{fig:SEM} for $N=5$). The ARB process used in the present study obviously affects the sequence and interaction of shear banding and texture evolution, leading to interesting deviations from the conventional, well-established model scenario described above. 

To put our observations into proper perspective, we note that the experimental conditions under which shear bands and texture development were observed in the present study differ in terms of several key aspects from previously published studies:
\begin{enumerate}[(i)]
	\item The CuZn15 alloy only exhibits low amounts of mechanical twinning compared to 70/30 brass investigated elsewhere, \cite{Jia12,Pasebani10}. 
	\item The ARB passes were performed at elevated temperatures (473~K) to improve the quality of the bonding. Together with additional heat introduced by the rolling itself, the increased temperatures promote thermally activated dislocation slip over mechanical twinning. In addition, these temperatures may even be enough to initiate recovery processes during ARB.
	\item In comparison to conventional rolling, the ARB process introduces large amounts of additional shear strain in the bulk of the material \cite{Lee02}.\label{redshear}
\end{enumerate}

All of these aspects can affect both the mechanism of shear band initiation and brass-type texture formation. A direct consequence of (\ref{redshear}) is the accelerated grain refinement during ARB compared to conventional rolling \cite{Huang03}. The smaller grain size during early stages of ARB processing may also influence the evolution of texture. Gu et al. have recently reported an unusual brass-type rolling texture in UFG copper after equal channel angular pressing followed by rolling \cite{Gu14}. They attributed this texture to the activation of {\{11\={1}\}<112>} partial slip, which is only activated in copper in the UFG regime without significant twinning. The SFE in CuZn15 (Grace and Inman estimated $23 mJ/m^2$ for CuZn16 \cite{Grace70}) is considerably lower than in copper ($78 mJ/m^2$ \cite{Loretto65}) which accelerates grain refinement during ARB \cite{Jamaati13}. Thus, the critical grain size to activate {\{11\={1}\}<112>} partial slip is likely to be achieved already at lower total plastic strains compared to pure copper. Finally, dynamic recrystallization can play a complex role in determining texture changes \cite{Kocks00} and the corresponding microstructural processes may be a key factor promoting brass texture formation prior to shear banding. 

To summarize, we have documented the development of a strong brass-type texture in CuZn15 during ARB prior to the onset of shear banding. Only little mechanical twinning was observed. The large additional shear strains introduced into the material during ARB, the resulting UFG microstructures with very fine grains, the elevated temperatures used in our study, and a SFE in between those of the baseline materials copper and 70/30 brass provide experimental conditions where the brass texture can develop already relatively early, while shear banding is observed only later. 

\section*{Acknowledgements}
We would like to thank Paul Fritzsche for help with the ARB experiments and Marc Pügner for help with the XRD texture measurements. 

\section*{References}
\bibliography{ARB}

\begin{thebibliography}{29}
\expandafter\ifx\csname natexlab\endcsname\relax\def\natexlab#1{#1}\fi
\providecommand{\bibinfo}[2]{#2}
\ifx\xfnm\relax \def\xfnm[#1]{\unskip,\space#1}\fi
\bibitem[{Hu et~al.(1952)Hu, Sperry, and Beck}]{Hu52}
\bibinfo{author}{H.~Hu}, \bibinfo{author}{P.~R. Sperry}, \bibinfo{author}{P.~A.
  Beck}, \bibinfo{journal}{JOM} \bibinfo{volume}{4} (\bibinfo{year}{1952})
  \bibinfo{pages}{76--81}.
\bibitem[{Wassermann(1963)}]{Wassermann63}
\bibinfo{author}{G.~Wassermann}, \bibinfo{journal}{Z. Metallkd.}
  \bibinfo{volume}{54} (\bibinfo{year}{1963}) \bibinfo{pages}{61}.
\bibitem[{Duggan et~al.(1978)Duggan, Hatherly, Hutchinson, and
  Wakefield}]{Duggan78}
\bibinfo{author}{B.~J. Duggan}, \bibinfo{author}{M.~Hatherly},
  \bibinfo{author}{W.~B. Hutchinson}, \bibinfo{author}{P.~T. Wakefield},
  \bibinfo{journal}{Metal Science} \bibinfo{volume}{12} (\bibinfo{year}{1978})
  \bibinfo{pages}{343--351}.
\bibitem[{Hirsch and Lücke(1988)}]{Hirsch88}
\bibinfo{author}{J.~Hirsch}, \bibinfo{author}{K.~Lücke},
  \bibinfo{journal}{Acta Metallurgica} \bibinfo{volume}{36}
  (\bibinfo{year}{1988}) \bibinfo{pages}{2863 -- 2882}.
\bibitem[{Leffers and Bilde-S{\o}rensen(1990)}]{Leffers90}
\bibinfo{author}{T.~Leffers}, \bibinfo{author}{J.~Bilde-S{\o}rensen},
  \bibinfo{journal}{Acta Metallurgica et Materialia} \bibinfo{volume}{38}
  (\bibinfo{year}{1990}) \bibinfo{pages}{1917--1926}.
\bibitem[{Leffers and Ray(2009)}]{Leffers09}
\bibinfo{author}{T.~Leffers}, \bibinfo{author}{R.~Ray},
  \bibinfo{journal}{Progress in Materials Science} \bibinfo{volume}{54}
  (\bibinfo{year}{2009}) \bibinfo{pages}{351 -- 396}.
\bibitem[{Paul et~al.(2002)Paul, Driver, and Jasieński}]{Paul02}
\bibinfo{author}{H.~Paul}, \bibinfo{author}{J.~Driver},
  \bibinfo{author}{Z.~Jasieński}, \bibinfo{journal}{Acta Materialia}
  \bibinfo{volume}{50} (\bibinfo{year}{2002}) \bibinfo{pages}{815 -- 830}.
\bibitem[{Paul et~al.(2003)Paul, Driver, Maurice, and Jasie{\'{n}}ski}]{Paul03}
\bibinfo{author}{H.~Paul}, \bibinfo{author}{J.~Driver},
  \bibinfo{author}{C.~Maurice}, \bibinfo{author}{Z.~Jasie{\'{n}}ski},
  \bibinfo{journal}{Materials Science and Engineering: A} \bibinfo{volume}{359}
  (\bibinfo{year}{2003}) \bibinfo{pages}{178--191}.
\bibitem[{Paul et~al.(2007)Paul, Driver, Maurice, and Piątkowski}]{Paul07}
\bibinfo{author}{H.~Paul}, \bibinfo{author}{J.~Driver},
  \bibinfo{author}{C.~Maurice}, \bibinfo{author}{A.~Piątkowski},
  \bibinfo{journal}{Acta Materialia} \bibinfo{volume}{55}
  (\bibinfo{year}{2007}) \bibinfo{pages}{575 -- 588}.
\bibitem[{El-Danaf et~al.(2000)El-Danaf, Kalidindi, Doherty, and
  Necker}]{ElDanaf00}
\bibinfo{author}{E.~El-Danaf}, \bibinfo{author}{S.~Kalidindi},
  \bibinfo{author}{R.~Doherty}, \bibinfo{author}{C.~Necker},
  \bibinfo{journal}{Acta Materialia} \bibinfo{volume}{48}
  (\bibinfo{year}{2000}) \bibinfo{pages}{2665--2673}.
\bibitem[{Jia et~al.(2012{\natexlab{a}})Jia, Roters, Eisenlohr, Kords, and
  Raabe}]{Jia12}
\bibinfo{author}{N.~Jia}, \bibinfo{author}{F.~Roters},
  \bibinfo{author}{P.~Eisenlohr}, \bibinfo{author}{C.~Kords},
  \bibinfo{author}{D.~Raabe}, \bibinfo{journal}{Acta Materialia}
  \bibinfo{volume}{60} (\bibinfo{year}{2012}{\natexlab{a}})
  \bibinfo{pages}{1099--1115}.
\bibitem[{Jia et~al.(2012{\natexlab{b}})Jia, Eisenlohr, Roters, Raabe, and
  Zhao}]{Jia12a}
\bibinfo{author}{N.~Jia}, \bibinfo{author}{P.~Eisenlohr},
  \bibinfo{author}{F.~Roters}, \bibinfo{author}{D.~Raabe},
  \bibinfo{author}{X.~Zhao}, \bibinfo{journal}{Acta Materialia}
  \bibinfo{volume}{60} (\bibinfo{year}{2012}{\natexlab{b}})
  \bibinfo{pages}{3415--3434}.
\bibitem[{Saito et~al.(1999)Saito, Utsunomiya, Tsuji, and Sakai}]{Saito99}
\bibinfo{author}{Y.~Saito}, \bibinfo{author}{H.~Utsunomiya},
  \bibinfo{author}{N.~Tsuji}, \bibinfo{author}{T.~Sakai},
  \bibinfo{journal}{Acta Materialia} \bibinfo{volume}{47}
  (\bibinfo{year}{1999}) \bibinfo{pages}{579 -- 583}.
\bibitem[{Huang et~al.(2003)Huang, Tsuji, Hansen, and Minamino}]{Huang03}
\bibinfo{author}{X.~Huang}, \bibinfo{author}{N.~Tsuji},
  \bibinfo{author}{N.~Hansen}, \bibinfo{author}{Y.~Minamino},
  \bibinfo{journal}{Materials Science and Engineering: A} \bibinfo{volume}{340}
  (\bibinfo{year}{2003}) \bibinfo{pages}{265 -- 271}.
\bibitem[{Pasebani et~al.(2010)Pasebani, Toroghinejad, Hosseini, and
  Szpunar}]{Pasebani10}
\bibinfo{author}{S.~Pasebani}, \bibinfo{author}{M.~R. Toroghinejad},
  \bibinfo{author}{M.~Hosseini}, \bibinfo{author}{J.~Szpunar},
  \bibinfo{journal}{Materials Science and Engineering: A} \bibinfo{volume}{527}
  (\bibinfo{year}{2010}) \bibinfo{pages}{2050--2056}.
\bibitem[{Pasebani and Toroghinejad(2010)}]{Pasebani10a}
\bibinfo{author}{S.~Pasebani}, \bibinfo{author}{M.~R. Toroghinejad},
  \bibinfo{journal}{Materials Science and Engineering: A} \bibinfo{volume}{527}
  (\bibinfo{year}{2010}) \bibinfo{pages}{491--497}.
\bibitem[{Chen et~al.(2009)Chen, Shi, Chen, Zhou, Wang, and Luo}]{Chen09}
\bibinfo{author}{L.~Chen}, \bibinfo{author}{Q.~Shi}, \bibinfo{author}{D.~Chen},
  \bibinfo{author}{S.~Zhou}, \bibinfo{author}{J.~Wang},
  \bibinfo{author}{X.~Luo}, \bibinfo{journal}{Materials Science and
  Engineering: A} \bibinfo{volume}{508} (\bibinfo{year}{2009})
  \bibinfo{pages}{37 -- 42}.
\bibitem[{Shaarbaf and Toroghinejad(2009)}]{Shaarbaf09}
\bibinfo{author}{M.~Shaarbaf}, \bibinfo{author}{M.~R. Toroghinejad},
  \bibinfo{journal}{Metallurgical and Materials Transactions A}
  \bibinfo{volume}{40} (\bibinfo{year}{2009}) \bibinfo{pages}{1693--1700}.
\bibitem[{Leffers and Jensen(1988)}]{Leffers88}
\bibinfo{author}{T.~Leffers}, \bibinfo{author}{D.~J. Jensen},
  \bibinfo{journal}{Textures and Microstructures} \bibinfo{volume}{8}
  (\bibinfo{year}{1988}) \bibinfo{pages}{467--480}.
\bibitem[{Leffers(1993)}]{Leffers93}
\bibinfo{author}{T.~Leffers}, \bibinfo{journal}{Textures and Microstructures}
  \bibinfo{volume}{22} (\bibinfo{year}{1993}) \bibinfo{pages}{53--58}.
\bibitem[{Christoffersen and Leffers(1997)}]{Christoffersen97}
\bibinfo{author}{H.~Christoffersen}, \bibinfo{author}{T.~Leffers},
  \bibinfo{journal}{Scripta Materialia} \bibinfo{volume}{37}
  (\bibinfo{year}{1997}) \bibinfo{pages}{1429--1434}.
\bibitem[{Grace and Inman(1970)}]{Grace70}
\bibinfo{author}{F.~Grace}, \bibinfo{author}{M.~Inman},
  \bibinfo{journal}{Metallography} \bibinfo{volume}{3} (\bibinfo{year}{1970})
  \bibinfo{pages}{89 -- 98}.
\bibitem[{Bachmann et~al.(2010)Bachmann, Hielscher, and Schaeben}]{bachmann10}
\bibinfo{author}{F.~Bachmann}, \bibinfo{author}{R.~Hielscher},
  \bibinfo{author}{H.~Schaeben}, in: \bibinfo{booktitle}{Texture and Anisotropy
  of Polycrystals III}, volume \bibinfo{volume}{160} of
  \textit{\bibinfo{series}{Solid State Phenomena}}, \bibinfo{publisher}{Trans
  Tech Publications}, \bibinfo{year}{2010}, pp. \bibinfo{pages}{63--68}.
\bibitem[{Bunge(1969)}]{Bunge69}
\bibinfo{author}{H.~Bunge}, \bibinfo{title}{Mathematische Methoden der
  Texturanalyse}, \bibinfo{publisher}{Akademie-Verlag}, \bibinfo{year}{1969}.
\bibitem[{Lee et~al.(2002)Lee, Saito, Tsuji, Utsunomiya, and Sakai}]{Lee02}
\bibinfo{author}{S.~Lee}, \bibinfo{author}{Y.~Saito},
  \bibinfo{author}{N.~Tsuji}, \bibinfo{author}{H.~Utsunomiya},
  \bibinfo{author}{T.~Sakai}, \bibinfo{journal}{Scripta Materialia}
  \bibinfo{volume}{46} (\bibinfo{year}{2002}) \bibinfo{pages}{281--285}.
\bibitem[{Gu et~al.(2014)Gu, Toth, Zhang, and Hoffman}]{Gu14}
\bibinfo{author}{C.~Gu}, \bibinfo{author}{L.~Toth}, \bibinfo{author}{Y.~Zhang},
  \bibinfo{author}{M.~Hoffman}, \bibinfo{journal}{Scripta Materialia}
  \bibinfo{volume}{92} (\bibinfo{year}{2014}) \bibinfo{pages}{51 -- 54}.
\bibitem[{Loretto et~al.(1965)Loretto, Clarebrough, and Segall}]{Loretto65}
\bibinfo{author}{M.~H. Loretto}, \bibinfo{author}{L.~M. Clarebrough},
  \bibinfo{author}{R.~L. Segall}, \bibinfo{journal}{Philosophical Magazine}
  \bibinfo{volume}{11} (\bibinfo{year}{1965}) \bibinfo{pages}{459--465}.
\bibitem[{Jamaati et~al.(2013)Jamaati, Toroghinejad, and Edris}]{Jamaati13}
\bibinfo{author}{R.~Jamaati}, \bibinfo{author}{M.~R. Toroghinejad},
  \bibinfo{author}{H.~Edris}, \bibinfo{journal}{Materials Science and
  Engineering: A} \bibinfo{volume}{578} (\bibinfo{year}{2013})
  \bibinfo{pages}{191--196}.
\bibitem[{Kocks et~al.(2000)Kocks, Tom{\'e}, and Wenk}]{Kocks00}
\bibinfo{author}{U.~Kocks}, \bibinfo{author}{C.~Tom{\'e}},
  \bibinfo{author}{H.~Wenk}, \bibinfo{title}{Texture and Anisotropy: Preferred
  Orientations in Polycrystals and Their Effect on Materials Properties},
  \bibinfo{publisher}{Cambridge University Press}, \bibinfo{year}{2000}.

\end{thebibliography}

\end{document}